\documentclass[%
 reprint,%
 amsmath,amssymb,
 superscriptaddress
 longbibliography,
]{revtex4-2}
\usepackage[dvipdfmx]{graphicx} % to include pdf figure
\usepackage{dcolumn}
\usepackage{bm}% bold math
\usepackage[utf8]{inputenc}
\usepackage[T1]{fontenc}
\usepackage{color}
\usepackage{amsmath}
\usepackage{url}
\usepackage[colorlinks=true,allcolors=blue,breaklinks]{hyperref}
\usepackage[page]{appendix}

\usepackage{siunitx}

\def\ket#1{\left|{#1}\right\rangle}
\def\bra#1{\left\langle{#1}\right|}

\let\temp\epsilon
\let\epsilon\varepsilon
\let\varepsilon\temp
\let\temp\relax
\let\temp\phi
\let\phi\varphi
\let\varphi\temp
\let\temp\relax

\begin{document}
%%%%%%%%%%%%%%%%%%%%%%%%%%%%%%%%%%%%%%%%%%%%%%%%%% 
 \title{
 Bridging Quantum Computing and Nuclear Structure: Atomic Nuclei on a Trapped-Ion Quantum Computer
 }
  %%%%%%%%%%%%%%%%%%%%%%%%% 
  \author{Sota Yoshida}
  \email{syoshida@a.utsunomiya-u.ac.jp}
  \affiliation{School of Data Science and Management, Utsunomiya University, Mine, Utsunomiya, 321-8505, Japan}
  \affiliation{RIKEN Nishina Center for Accelerator-based Science, RIKEN, Wako 351-0198, Japan}
  %ORCID: 0000-0002-1342-1846
  %%%%%%%%%%%%%%%%%%%%%%%%% 
  \author{Takeshi Sato}
  \affiliation{Graduate School of Engineering, The University of Tokyo, 7-3-1 Hongo, Bunkyo-ku, Tokyo 113-8656, Japan}
  \affiliation{Photon Science Center, School of Engineering, The University of Tokyo, 7-3-1 Hongo, Bunkyo-ku, Tokyo 113-8656, Japan}
  \affiliation{Research Institute for Photon Science and Laser Technology, The University of Tokyo, 7-3-1 Hongo, Bunkyo-ku, Tokyo 113-0033, Japan}
  %ORCID: 0000-0001-8952-7885
  %%%%%%%%%%%%%%%%%%%%%%%%% 
  \author{Takumi Ogata}
  \affiliation{Graduate School of Engineering, The University of Tokyo, 7-3-1 Hongo, Bunkyo-ku, Tokyo 113-8656, Japan}
  %%%%%%%%%%%%%%%%%%%%%%%%% 
  \author{Masaaki Kimura}
  \affiliation{RIKEN Nishina Center for Accelerator-based Science, RIKEN, Wako 351-0198, Japan}
  \affiliation{Department of Physics / Quark Nuclear Science Institute, Graduate School of Science, The University of Tokyo, 7-3-1 Hongo, Bunkyo-ku, Tokyo 113-0033, Japan}

%%%%%%%%%%%%%%%%%%%%%%%%%%%%%%%%%%%%%%% 
\begin{abstract}
We demonstrate quantum simulations of strongly correlated nuclear many-body systems 
on the RIKEN-Quantinuum Reimei trapped-ion quantum computer, 
targeting ground states of oxygen, calcium, and nickel isotopes. 
By combining a hard-core-boson representation of the nuclear shell model with a pair-unitary coupled-cluster doubles ansatz, 
we achieve sub-percent relative error in the ground-state energies compared to noise-free statevector simulations. 
Our approach leverages symmetry-aware state preparation and particle-number post-selection 
to efficiently capture pairing correlations characteristic of systems with same-species nucleons.
These findings highlight the viability of high-fidelity trapped-ion platforms 
for nuclear physics applications and provide a foundation for scaling to more complex nuclear systems.
\end{abstract}
%%%%%%%%%%%%%%%%%%%%%%%%%%%%%%%%%%%%%%% 
\maketitle

%%%%%%%%%%%%%%%%%%%%%%%%%%%%%%%%%%% 
%%%%%% +++++ Intro +++++%%%%%%%%%%%
%%%%%%%%%%%%%%%%%%% %%%%%%%%%%%%%%%

\section{Introduction}
\par

% overview of quantum computing for quantum many-body systems
Quantum simulation of many-body systems is a leading application of quantum computing, with potential impact across quantum chemistry, condensed-matter physics, nuclear physics and so on.
While quantum algorithms and hardware have advanced rapidly, demonstrating clear advantage for interacting many-body problems remains an open challenge.
For example, a recent study~\cite{Yoshioka2024_npjQI} identified scenarios in which quantum devices can outperform classical algorithms for certain spin models.
Nonetheless, general claims of quantum advantage for many-body systems remain inconclusive and require careful, system-specific analysis.

% Quantum algorithms and current situation
On near-term devices, variational approaches such as
the Variational Quantum Eigensolver~\cite{Peruzzo_14NatCom, Bharti2022}
rely crucially on the choice of an ansatz that balances implementability and expressivity.
Common families of ansatze — including unitary coupled cluster (UCC) variants — can be
effective in some domains but often suffer from optimization difficulties,
called barren plateaus~\cite{VQE_BarrenPlateau2025},
or unfavorable scaling of parameters and gate depth as system size grows.
This situation has evoked a broader consideration of quantum algorithms
beyond noisy intermediate-scale quantum (NISQ) towards fault-tolerant quantum computing (FTQC),
leading to a term like Early Fault-Tolerant Quantum Computing (E-FTQC)~\cite{Katabarwa2024}.
In such situations, it is essential to broadly examine each component of quantum computation—namely,
state preparation (also simply called the ansatz), quantum algorithms,
and measurement strategies for evaluating the observable of interest—for each specific problem.

The first part, the choice of the ansatz, is particularly crucial.
This is because the success of most of quantum algorithms for quantum many-body problems
largely depends on the quality of the prepared quantum state,
i.e., the overlap between the prepared state and the target state.
In spin systems or chemical systems, various ansatze have been proposed, such as UCC-family,
and ones based on matrix product states or tensor network states more generally~\cite{Ran_PRA20,Felser_PRL21,Malz2024_MPS}.
Compared to other areas tackling quantum many-body problems,
nuclear physics studies have seen relatively limited applications of quantum computing,
despite several pioneering works~\cite{Deuteron_PRL18, LacroixPRL20, Spain_SciRepo2023},
and recent reviews~\cite{GRamos23, Savage24}.
To date, only a few ansatze have been explored for nuclear systems, almost exclusively
based on the unitary coupled cluster (UCC) family~\cite{Kiss_PRC22, Sarma_PRC2023, Spain_SciRepo2023}
with some variations such as Adapt-VQE~\cite{Grimsley2019}.
A seminal work investigated the applicability of the UCC + Adapt-VQE
to the nuclear shell model~\cite{Spain_SciRepo2023},
where the number of CNOT gates required to achieve a given accuracy
for shell-model calculations of oxygen and calcium isotopes was estimated.
Taking a look around existing literature, pioneering works have focused intensively
on pairing Hamiltonians~\cite{LacroixPRL20, Guzman_PRC22, PhysRevC.107.044308, np_PhysRevC.110.064320, ERGuzman2024_EPJA, LiuPLB85_102480, ZhangPLB869_139841}.
Building on these previous works on pairing Hamiltonians or pairing-like ansatze,
this work aims to explore more expressive and scalable ansatze suitable for simplified
representations of nuclear shell-model Hamiltonians, develop symmetry-aware post-selection techniques,
and evaluate their performance on quantum hardware across a range of nuclear systems.

Nuclear many-body systems are distinguished from other quantum many-body systems
by exceptionally strong correlations arising from the nature of nuclear forces,
posing a unique challenge for both classical and quantum computing.
Even in the low-energy regime, where nucleons are the only explicit baryonic degrees of freedom,
the problem is complicated by the fact that protons and neutrons---fermions of nearly equal
mass---interact through multiple channels in an intricate manner.
This results in a large number of non-negligible Hamiltonian terms and strong non-locality.
Furthermore, anisotropic components of the interaction, such as tensor forces~\cite{OtsukaPRL05}
and three-body forces~\cite{Hebeler21Rev}, play an essential role in describing the
rich phenomena of nuclear systems.
These features are generally difficult to capture with a simple wave-function ansatz.
Another manifestation of the complexity is entanglement entropy, which in nuclear systems
obeys a volume law rather than an area law~\cite{Gu2023},
hindering straightforward application of tensor networks.
% Extensions of tensor-network methods to such systems
% remain a promising direction~\cite{Bettaque24, Liu24, Lu2025}.

Consequently, most quantum-computing studies of nuclear structure have focused on small systems
or relied on classical simulators.
Reported errors in ground-state energy estimates on quantum hardware
range from $4 - 13 \%$ for ${}^{6}$Li~\cite{Kiss_PRC22}
and $3 - 13 \%$ for oxygen isotopes~\cite{Sarma_PRC2023}.
In our previous work~\cite{SY_PRC2024}, we achieved relative errors
below $0.1 \%$ for ${}^{6}$He and ${}^{18}$O,
and about $1\%$ for ${}^{42}$Ca on an IBM superconducting quantum device.
However, the target systems were limited to valence two-neutron configurations
and our method is not straightforwardly extendable to more general nuclear systems.
This motivates the present work to develop more general and scalable approaches.

In this work, we introduce a hard-core-boson (HCB) mapping of the nuclear shell model
combined with the pair-unitary coupled-cluster doubles (pUCCD) ansatz,
which efficiently captures strong pairing correlations between nucleons of the same species.
While pUCCD ansatze have been shown to be effective for chemical systems~\cite{Elfving_PRA2021, GoogleNP2023, Zhao2023},
applications of such pair-based ansatze to nuclear systems have been limited a few pioneering studies:
\cite{np_PhysRevC.110.064320, ZhangPLB869_139841, CostaPLB872_140042}.
This work extends these previous studies to shell-model Hamiltonians,
and reports the first hardware implementation of HCB and pUCCD for nuclear systems.
Implemented on the high-fidelity trapped-ion processor RIKEN-Quantinuum Reimei,
and augmented by symmetry-aware ansatze and post-selection techniques,
our approach yields ground-state energies for oxygen, calcium, and nickel isotopes
that agree with noise-free statevector simulations within sub-percent relative error.
These results establish a new experimental benchmark for quantum simulations of 
strongly correlated nuclear systems and demonstrate the near-term potential
of trapped-ion platforms for nuclear-structure calculations.

This paper is organized as follows.
In Sec.~\ref{sec:methods}, we describe the methods used in this work, including the Hamiltonian mapping,
the adopted ansatz, and the measurement strategies.
In Sec.~\ref{sec:Results}, we present the results for oxygen, calcium, and nickel isotopes,
noise-free simulations and hardware experiments.
Finally, we summarize our findings and discuss future outlooks in Sec.~\ref{sec:Summary}.

%
%%%%%%%%%%%%%%%%%%%%%%%%%%%%%%%%%%%
%%%%%%% +++++ Methods +++++ %%%%%%%
%%%%%%%%%%%%%%%%%%%%%%%%%%%%%%%%%%%
% 
\section{Methods \label{sec:methods}}

\subsection{Target systems and Hamiltonian mapping \label{subsec:HamilBasis}}

General shell-model Hamiltonians with up to two-body interactions can be expressed as 
\begin{align}
  \hat{H} & = \sum_{i} h_{ij} a^\dagger_i a_j 
  + \frac{1}{4} \sum_{ijkl} V_{ijkl} a^\dagger_i a^\dagger_j a_l a_k,
  \label{eq:SMhamiltonian}
\end{align}
where $ i $ denotes the single-particle state characterized by the quantum numbers
$ \left\{ n, l, j, j_z, t_z \right\} $.
Here, $n$ is the principal quantum number, $l$ the orbital angular momentum,
$j$ the total angular momentum, $j_z$ its projection onto the $z$-axis,
and $t_z$ the isospin projection.

For systems within single major shell, as in this work,
the one-body term reduces to the diagonal form $ \epsilon_i \equiv h_{ii} $,
called the single-particle energies (SPEs).
The fermionic creation and annihilation operators, $a^\dagger_i$ and $a_i$,
act on these single-particle states, while $ V_{ijkl} $ 
denotes the two-body matrix elements (TBMEs).

These SPEs and TBMEs can be either derived from the realistic nucleon-nucleon interactions
or from the phenomenological adjustments to reproduce experimental data for binding energies, excitation energies, and so on.
The target nuclei, model space and adopted interactions in this work are summarized in Table~\ref{tab:interactions}.
The methods in this work can be applied to any other interactions including microscopically derived ones
as far as one considers similar systems with same-species nucleons under the methods described below.
Diagonalization with these interactions can be reproduced using publicly available codes such as
KSHELL~\cite{KSHELL1, *KSHELL2} or 
NuclearToolkit.jl~\cite{ NuclearToolkit.jl, *Repo_NuclearToolkit.jl}.

\begin{table}[b]
  \begin{ruledtabular}
  \caption{Target nuclei, interactions, and model spaces.
  \label{tab:interactions}
  }
  \begin{tabular}{ccc}
    Nuclide & Interaction & Model space  \\
    \hline
    ${}^{18-26}$O & USDB~\cite{USDB} & $sd$ ($1s_{1/2}, 0d_{3/2}, 0d_{5/2}$) \\
    ${}^{42-58}$Ca & GXPF1A~\cite{GXPF1A} & $pf$ ($1p_{1/2}, 1p_{3/2}, 0f_{5/2}, 0f_{7/2}$) \\
    ${}^{58-76}$Ni & JUN45~\cite{JUN45} & $jj45$ ($1p_{1/2}, 1p_{3/2}, 0f_{5/2}, 0g_{9/2}$) \\
  \end{tabular}
  \end{ruledtabular}
\end{table}

Once the Hamiltonian is defined, one can diagonalize the Hamiltonian in a given model space
by means of variants of Configuration Interaction (CI) methods,
known as the nuclear shell model in nuclear physics community.
In this work, full configuration interaction (FCI) refers to exact diagonalization
within the selected valence space, following the convention of quantum chemistry
(rather than the no-core shell model definition often used in nuclear physics).

A useful truncation of CI is the doubly occupied configuration interaction (DOCI) method,
equivalent to restricting the shell-model space to zero-seniority configurations,
i.e., considering only time-reversed nucleon pairs as active degrees of freedom.
In even-neutron systems, DOCI often provides a good approximation to FCI 
and is exact for two-valence-neutron $J=0$ states.
This DOCI method has been also used in nuclear physics and proven to be a good approximation
to even-number neutron systems~\cite{Volya2001}.
Thus, DOCI serves as an important benchmark in this work.

Another viable method for approximating FCI and/or DOCI is the use of pair coupled cluster methods
and their extensions~\cite{Henderson_PRC2014, Henderson_JChemPhys2015, Qiu2019, Elfving_PRA2021, GoogleNP2023, Zhao2023}.
Whereas the methods have been successfully applied to pairing Hamiltonians (also known as BCS Hamiltonians) in the nuclear physics context,
their applications to more general shell-model Hamiltonians remain limited and warrant further exploration.

In the case of electronic structure calculations, the natural assumption on the 
{\it pair} degree of freedom is to consider the spin up and spin down electrons as a pair.
Detailed derivations of such zero-seniority Hamiltonian for electronic structure calculations 
can be found in e.g.~\cite{Henderson_JChemPhys2015}.
For nuclear many-body methods one has more freedom involved in the system,
so one can consider the pairs of nucleons with different spins and/or different orbitals.
For this reason, we will detail the way to introduce the paired form of a nuclear shell-model Hamiltonian.

The hard-core-boson (HCB) representation provides a natural way to introduce paired degrees of freedom
by folding each time-reversed single-particle pair $(i,\bar{i})$ into a
single boson-like site that can be occupied or empty but forbids multiple occupancy.
Similar mappings have been employed in various contexts within quantum chemistry and nuclear physics
\cite{Khamoshi2020_AGP, Henderson2020_JChemPhys, Elfving_PRA2021, Guzman_PRC22, np_PhysRevC.110.064320, SY_PRC2024,ZhangPLB869_139841, CostaPLB872_140042}.
Depending on whether one emphasizes the underlying degrees of freedom 
or the commutation relations of the operators, this construction is referred
to by various names—geminal, hard-core-boson~(HCB),
antisymmetrized geminal power~(AGP), seniority-zero,
quasi-particle representations, etc.—but the underlying idea is similar.

\begin{figure}%[htbp]
  \centering
  \includegraphics[width=\linewidth]{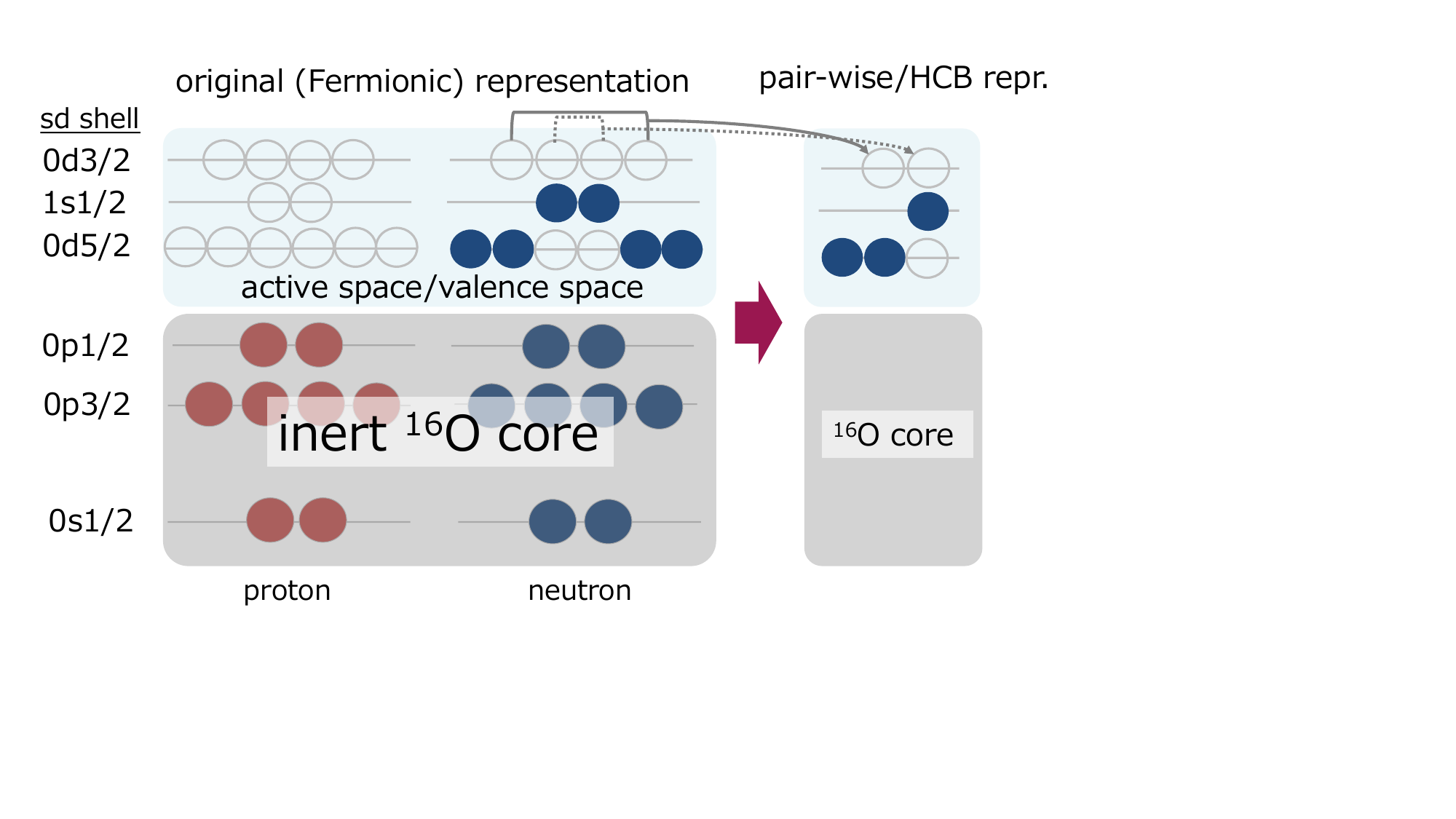}
  \caption{
    Schematics of the hard-core boson (HCB) representation.
    This example shows a configuration of ${}^{22}$O nucleus on top of the inert core ${}^{16}$O.
    The active (valence) space consists of $1s_{1/2}, 0d_{3/2}$, and $0d_{5/2}$ single-particle states.
    \label{fig:sketch_HCB}
  }
\end{figure}

Here, we summarize the derivation of the HCB representation for nuclear shell-model Hamiltonians.
Figure~\ref{fig:sketch_HCB} illustrates this mapping for ${}^{22}$O on top of the
${}^{16}$O inert core: the valence single-particle states are grouped into pair orbitals,
which reduces the number of active degrees of freedom (and hence qubits).
The single-particle states in the original fermionic representation
are specified by the quantum numbers $ \{ n, l, j, j_z, t_z \} $,
leading to 12 single-particle states for both protons and neutrons in the $sd$ shell.
In the HCB representation, these 12 single-particle states form 6 pair orbitals for neutrons,
each represented by a qubit.
Focusing on even-number neutron systems and terms involving time-reversed pairs,
the shell-model Hamiltonian can be expressed as
\begin{align}
  H & =  \sum_{i<0} \epsilon_i ( a^\dagger_i a_i + a^\dagger_{\bar{i}} a_{\bar{i}} ) 
  + \sum_{i<0} V_{i \bar{i} i \bar{i}} a^\dagger_i a^\dagger_{\bar{i}} a_{\bar{i}} a_i \nonumber \\
  & + \sum_{i \neq j <0} V_{i\bar{i}j\bar{j}} a^\dagger_i a^\dagger_{\bar{i}} a_{\bar{j}} a_j
  + \frac{1}{4} \sum_{i, j\notin \{i, \bar{i}\}} V_{ijij} n_i n_j ,
  \label{eq:Hamil_fermionic}
\end{align}
where $i<0$ denotes single-particle states with negative $j_z$ values,
and $\bar{i}$ is the time-reversal partner of the single-particle state $i$.
The last term, absent in the valence two-neutron systems studied previously~\cite{SY_PRC2024},
can be regarded as a mean-field contribution from other occupied pairs.
Then, the Hamiltonian can be expressed in terms of the paired operators defined below:
  \begin{align}
    A^{\dag}_{\tilde{i}}
    & =
      a^{\dag}_i a^{\dag}_{\bar{i}},  \label{eq:opAdag}\\
    A_{\tilde{i}}
    & =
      a_{\bar{i}} a_i, \label{eq:opA} \\
    N_{\tilde{i}} & = \frac{1}{2} (n_i + n_{\bar{i}}) = \frac{1}{2} \left( 
      a^{\dag}_i a_i + a^{\dag}_{\bar{i}} a_{\bar{i}}  \right), \label{eq:opN}
  \end{align}
where $A^{\dag}_{\tilde{i}}$ and $A_{\tilde{i}}$ are the pair creation and annihilation operators,
and $N_{\tilde{i}}$ is the pair occupation number operator,
$\tilde{i}$ denotes the pair index corresponding to the single-particle states $i$ and $\bar{i}$.
Then, the Hamiltonian in Eq.~\eqref{eq:Hamil_fermionic} can be rewritten as
\begin{align}
  H_\mathrm{HCB}  =  & \sum_{\tilde{i}} (\epsilon_{\tilde{i}}+ V^p_{\tilde{i} \tilde{i}}) N_{\tilde{i}} \nonumber \\
&  + \sum_{\tilde{i} \neq \tilde{j}} V^p_{\tilde{i} \tilde{j}}A^\dagger_{\tilde{i}} A_{\tilde{j}}
  + \sum_{\tilde{i}\neq \tilde{j}} V^m_{\tilde{i} \tilde{j}} N_{\tilde{i}} N_{\tilde{j}}, \\
  \epsilon_{\tilde{i}} \equiv & 2 \epsilon_i = \epsilon_i + \epsilon_{\bar{i}}, \\
  V^p_{\tilde{i} \tilde{j}} \equiv & V_{i \bar{i} j \bar{j}}, \\
  V^m_{\tilde{i} \tilde{j}} \equiv & V_{ijij} + V_{i\bar{j}i\bar{j}} + V_{\bar{i}j\bar{i}j} + V_{\bar{i}\bar{j}\bar{i}\bar{j}},
  \label{eq:Hamil_pairwise}
\end{align}
where $V^p$ and $V^m$ are introduced to emphasize the pairing and monopole nature of the interactions, respectively.
The factor $1/2$ in Eq.~\eqref{eq:opN} is sometimes omitted (as in our earlier work~\cite{SY_PRC2024}),
but is retained here to interpret $N_{\tilde{i}}$ as the occupation number of a pair degree of freedom.
More details on this derivation are provided in the Supplemental Material~\cite{Supple}.

\begin{figure*}
  \centering
  \includegraphics[width=0.93\linewidth]{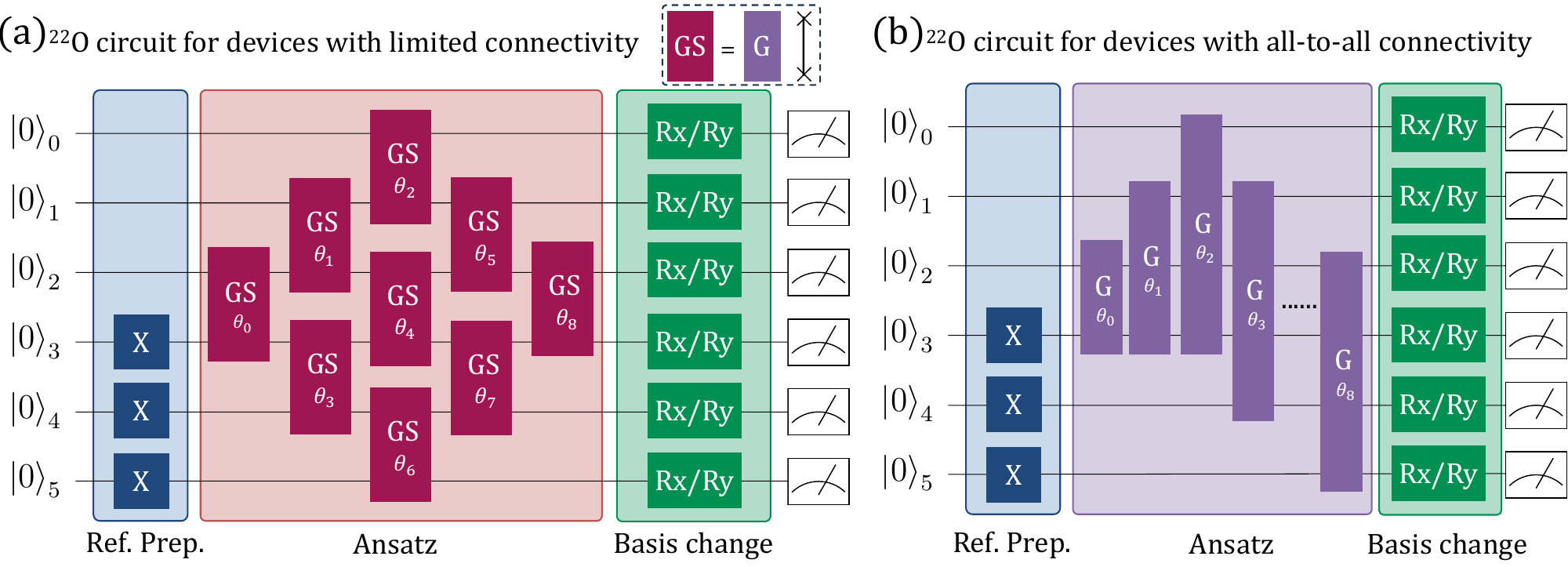}
  \caption{
  Sketched quantum circuits for preparing the pUCCD ansatz taking ${}^{22}$O as an example.
  Original single-particle states in the $sd$ shell consist of twelve states for neutrons,
  which are now represented by six qubits in the HCB representation.
  (a) Circuit layout suitable for devices with heavy-hexagon connectivity.
  (b) Circuit layout for all-to-all connectivity.
  First, the reference state is prepared by applying $X$ gates to qubits.
  Then, Givens rotations are applied to take into account particle-hole excitations of nucleon pairs.
  Single-qubit rotations are applied as needed prior to measuring Hamiltonian expectation values.
  \label{fig:Sketch_Circuits}
 }
\end{figure*}

The paired-neutron operators satisfy bosonic commutation relations,
as detailed in e.g.~\cite{Henderson_JChemPhys2015, Elfving_PRA2021, SY_PRC2024},
but multiple occupancy of the same pair is forbidden, leading to the term hard-core boson.
A key advantage of the HCB representation—beyond the qubit reduction—is the elimination of long strings of Pauli-$Z$ operators.
From Jordan-Wigner mappings, Pauli-$Z$ terms appear to maintain the anti-commutation relations of the fermionic operators.
In some literature, these Pauli-$Z$ terms are omitted on purpose,
which is referred to as qubit excitation in Refs.~\cite{Yordanov_PRA2020,Yordanov2021},
as a hardware-efficient mapping.
However, in the HCB representation, those Pauli-$Z$ terms naturally cancel out.
Hence, one can map the Hamiltonian to the Pauli operators into simple forms as
\begin{align}
H_\mathrm{HCB;q}
& =  
\sum_{i} \frac{\epsilon_{i} + V_{i i}}{2} (I_{i} - Z_{i})
+ \frac{1}{4} \sum_{i \neq j} V_{i j} (X_{i} X_{j} + Y_{i} Y_{j})   \nonumber \\
& + \frac{1}{4} \sum_{i \neq j} V^m_{i j} (I_{i} - Z_{i})(I_{j} - Z_{j}),
  \label{eq:Hamil_mapped}
\end{align}
where $i$ denotes the pair index of single-particle states in the valence space specified by
$n$ (principal quantum number), $l$ (orbital angular momentum), $j$ (total angular momentum),
$|j_z|$ (absolute value of the angular momentum projection on the $z$-axis), and $t_z$ (isospin projection) quanta.

All terms of the Hamiltonian Eq.~\eqref{eq:Hamil_mapped} except the second are diagonal in the computational basis
and can be measured directly by measuring the expectation values of the Pauli-$Z$ operators.
One needs additional circuits to measure the non-diagonal terms $XX+YY$,
which will be discussed later in Sec.~\ref{subsec:Measurement}.

\subsection{ansatze \label{subsec:ansatz}}

The ansatz, circuit for state preparation, plays a central role in quantum algorithms both in the NISQ and FTQC regimes.
In practice, the ansatz must satisfy two requirements:
(1) it should be implementable on available quantum hardware, and
(2) it should be expressive enough to approximate, e.g., the true ground state with high fidelity.

In this work, we employ the pair unitary coupled cluster doubles (pUCCD) ansatz  within the HCB representation,
which is particularly well-suited for systems with an even number of valence neutrons.
The pUCCD ansatz is a quantum version of the pair coupled cluster doubles method.
A seminal work on the pUCCD ansatz in the context of quantum computing
is found in Ref.~\cite{Elfving_PRA2021},
where strategies for measurement and post-selection were also proposed.
In chemistry, pUCCD has been successfully applied to systems of paired electrons
with spin up and spin down~\cite{Elfving_PRA2021, GoogleNP2023, Zhao2023}.
For nuclear systems, the pUCCD ansatz has been considered in a few theoretical studies~\cite{np_PhysRevC.110.064320},
but to our knowledge, no experimental implementation has been reported so far.

The pUCCD ansatz is expressed as
\begin{align}
\ket{\psi_\mathrm{pUCCD}} 
& = \exp{(T-T^\dag)} \ket{\mathrm{ref.}} \equiv U \ket{\mathrm{ref.}},
\end{align}
where $T$ is the cluster operator and $\ket{\mathrm{ref.}}$ is a reference state.
For paired-neutron excitations, the unitary operator takes the form
\begin{align}
U & = \exp{ \left[  \sum_{ph} t^p_h (A^\dagger_pA_h - A^\dagger_h A_p) \right] },
\end{align}
with $t^p_h$ the cluster amplitudes,
and $A^\dagger_p, A_h$ the pair creation and annihilation operators,
Eqs.~\eqref{eq:opAdag} and \eqref{eq:opA}, respectively.
The indices $p$ and $h$ denote the particle and hole states, respectively.
Mapping to Pauli operators yields
\begin{align}
U_\mathrm{qubit} 
& = \exp{ \left[ \sum_{ph} \frac{i t^p_h}{2} (X_p Y_h - Y_p X_h) \right] }, \\
& \approx \prod_{ph} \exp{ \left[ \frac{i t^p_h}{2} (X_p Y_h - Y_p X_h) \right] },
\label{eq:U_pUCCD}
\end{align}
where the last expression corresponds to a first-order Trotter approximation
with a single step, which we adopt in this work.
Each terms in Eq.~\eqref{eq:U_pUCCD} can be implemented as a Givens rotation gate acting on qubits $p$ and $h$:
\begin{align}
G(2\theta) & = \begin{pmatrix}
1 & 0 & 0 & 0 \\
0 & \cos{\theta} & -\sin{\theta} & 0 \\
0 & \sin{\theta} & \cos{\theta} & 0 \\
0 & 0 & 0 & 1
\end{pmatrix}.
\label{eq:Givens}
\end{align}

Thus, the pUCCD ansatz can be implemented as a sequence of Givens rotations
on top of a reference state, e.g., initially prepared by applying $X$ gates to the qubits corresponding to occupied pair orbitals.
Fig.~\ref{fig:Sketch_Circuits} illustrates two circuit layouts for preparing the pUCCD ansatz,
taking ${}^{22}$O nucleus as an example.
(a) pUCCD(GS): optimized for devices with limited connectivity (e.g., IBM heavy-hexagon architectures),
using Givens rotations followed by SWAP gates (GS)~\cite{Elfving_PRA2021}.
The basic idea behind the circuit is detailed in Ref.~\cite{Kivlichan2018}.
(b) pUCCD(G): one realization for all-to-all connectivity (e.g., trapped-ion devices) using only Givens rotations (G).
In both cases, the number of parameters is the same as the number of particle-hole pairs.

However, the circuit depth and number of two-qubit gates slightly differ.
While pUCCD(GS) has a nearly unique circuit layout, pUCCD(G) admits many realizations depending on the choice
of initial configuration and ordering of Givens rotations.
We will therefore present simulator results for pUCCD(G) as distributions over multiple circuit realizations in Sec.~\ref{sec:Results}.

In what follows, we use a single deterministic realization for both pUCCD(GS) and pUCCD(G) unless otherwise noted.
The qubits are aligned by ascending order of $j$, total angular momentum of the corresponding single-particle states.
This is not the optimal lowest-filling configuration; however,
we verified that initializing qubits strictly in ascending SPE order
is not necessarily superior. The present ordering is therefore a simple 
and reasonable choice among equivalent alternatives.
One can further confirm that this choice does not give an {\it outlier} result
by comparing with multiple random realizations, as will be shown later in Sec.~\ref{sec:Results}.

As a related study, it should be noted that a work by Sarma et al.~\cite{Sarma_PRC2023}
considered oxygen isotopes in the $sd$ shell on a trapped-ion quantum computer, IonQ Aria,
using a variant of unitary coupled cluster doubles (UCCD) ansatz.
Their UCCD ansatz was applied only to time-reversal nucleon pairs (denoted UCCD($\nu=0$) in the following where $\nu$ is the seniority).
Although formally equivalent to DOCI, their implementation yielded ground-state errors of $3-13\%$
relative to FCI, depending on the isotope.
One limitation arises from the scaling of circuit parameters:
the number of Givens rotations to express the excitations/de-excitations of nucleon pairs
grows rapidly with system size, making direct application to heavier isotopes impractical.

By contrast, the pUCCD ansatz with a single Trotter step scales more favorably, as 
\begin{align}
N_\mathrm{param} & = N_\mathrm{occ.} (N_q - N_\mathrm{occ.}),
\end{align}
where $N_\mathrm{occ.}$ is the number of occupied pairs and 
$N_q$ the number of qubits in the HCB representation.
This scaling is significantly milder than that of UCCD($\nu=0$), as summarized in Table~\ref{tab:ansatz_params}.
Note that the FCI dimension is also shown in the table for reference,
and these show the number of configurations in the so-called
M-scheme, i.e. the basis states are labeled by the sum of
the $j_z$ values of the single-particle states.
For even number of neutrons, all the states with different total angular momentum $J$
are included in the $M=0$ subspace leading to the FCI dimension shown in the table.

\begin{table}%[b]
  \begin{ruledtabular}
  \caption{Comparison of Hilbert-space dimension and ansatz parameter counts.
    For each nucleus we list: the number of qubits $N_q$ in both Fermionic and HCB representations,
    the full configuration-interaction (FCI) dimension in the $M=0$ sector,
    the number of variational parameters required by UCCD($\nu=0$) (unitary coupled-cluster doubles
    restricted to zero-seniority/paired configurations), and the number of parameters for the
    pUCCD ansatz implemented with a single Trotter step.
  \label{tab:ansatz_params}}
  \begin{tabular}{lcrrr}
            & $N_q$         & \multicolumn{3}{c}{Dimensions}   \\
    \cline{3-5}
    Nucleus & Fermionic/HCB & FCI & UCCD($\nu=0$) & pUCCD \\
    \hline
    $^{18}$O, $^{26}$O   & 12/6  &14& 5   & 5  \\
    $^{20}$O, $^{24}$O   &       &81& 14  & 8  \\
    $^{22}$O             &       &142& 19  & 9  \\ 
    $^{42}$Ca, $^{58}$Ca & 20/10 &30& 9   & 9  \\
    $^{44}$Ca, $^{56}$Ca &       &565& 44  & 16 \\
    $^{46}$Ca, $^{54}$Ca &       &3,952& 119 & 21 \\
    $^{48}$Ca, $^{52}$Ca &       &12,022& 209 & 24 \\
    $^{50}$Ca            &       &17,276& 251 & 25 \\
    $^{58}$Ni, $^{76}$Ni & 22/11 &19&  10 & 10 \\
    $^{60}$Ni, $^{74}$Ni &       &365&  54 & 18 \\
    $^{62}$Ni, $^{72}$Ni &       &3,103& 164 & 24 \\
    $^{64}$Ni, $^{70}$Ni &       &12,240& 329 & 28 \\
    $^{66}$Ni, $^{68}$Ni &       &23,884& 461 & 30 
  \end{tabular}
  \end{ruledtabular}
\end{table}

\subsection{Measurement of energy expectation values\label{subsec:Measurement}}

Once the ansatz circuit is prepared, the next task is to evaluate
the expectation value of the Hamiltonian.
First, we discuss our optimization approach; then we detail the measurement strategies.
In this work, the parameters of the pUCCD ansatze were optimized on statevector simulators
using the Adam optimizer with PennyLane~\cite{PennyLane}.
This is still an idealized setting, but good starting point to
evaluate the ansatz quality and measurement strategies, and to benchmark hardware results.
The strategy to optimize the ansatz circuit on a real device is left as a future issue.
Optimization directly on quantum hardware remains an open problem,
for which gradient-free sequential approaches~\cite{NFT_PRR2020} may offer a promising path.

Once the ansatz circuit is prepared, the next task is to evaluate
the expectation value of the Hamiltonian.
For noise-free simulations, one does not need to care about the measurement strategy,
but it is crucial point to be considered for noisy simulations and experiments on hardware.

Diagonal terms in the computational basis, i.e., operators involving $ \{I, Z_i, Z_i \otimes Z_j\} $,
can be obtained by measuring the ansatz circuit in the computational basis.
In this case, post-selection techniques is naturally applicable:
bitstrings violating the particle number of the target system are discarded.

In contrast, non-diagonal terms $ \{X_i \otimes X_j,  Y_i \otimes Y_j\} $ require additional circuits.
Both $XX$ and $YY$ terms are local, i.e., free from $Z$-strings,
and contribute equally to the energy expectation value.
Thus, all $X_pX_q$ terms can be evaluated by a single additional circuit,
ansatz followed by Hadamard gates on all qubits prior to measurement,
and $\langle Y_pY_q \rangle$ terms can be evaluated by doubling the $\langle X_pX_q \rangle$ results.
This approach will be referred to as the {\it Hadamard} method in the following.
While straightforward, this method does not preserve particle-number symmetry,
limiting the use of post-selection.
The other approach employed in this work is one to be referred to as {\it Basis rotations},
which involves diagonalizing $XX+YY$ terms in the computational basis prior to measurement~\cite{Google_2020, Elfving_PRA2021}:
\begin{align}
  \bra{\psi} & X_pX_q  + Y_pY_q \ket{\psi} \nonumber \\
  & = \bra{\psi} \mathcal{U}_{p,q}(\pi/4) (Z_p - Z_q) \mathcal{U}^\dagger_{p,q}(\pi/4) \ket{\psi},
\end{align}
where $\mathcal{U}_{p,q}(\theta)$ can be implemented as a Givens rotation $G(2\theta)$ on qubits $p$ and $q$.
In accordance with the discussion in Ref.~\cite{Google_2020},
the required terms can be grouped such that only $2 \lfloor N_q/2 \rfloor$ additional
circuits are needed to measure all the $XX+YY$ terms, where $N_q$ is the number of qubits.
Importantly, this strategy preserves particle-number symmetry, enabling effective post-selection.

Note that since we measure diagonal and off-diagonal contributions in separate circuits,
the summed estimate of the energy is not strictly variational even after post-selection.
However, in practice, we observed that the energy estimates
distributed around the noise-free values from statevector simulations,
hence non-variational nature does not pose significant issues in this work.

\subsection{Implementation of Givens rotation gates \label{subsec:app_Givens} }

To construct the pUCCD ansatz circuits (Fig.~\ref{fig:Sketch_Circuits})
and to perform basis rotations for diagonalizing the $XX+YY$ terms,
it is necessary to implement Givens rotation gates.
Several approaches are known in literature and we examined the following three methods:
\begin{enumerate}
  \item Magic-gate method (Ref.~\cite{Vatan_PRA04}) - implements Givens rotation using a sequence of "magic" gates.
  \item $\sqrt{\text{SWAP}}$-based method (Ref.~\cite{Kivlichan2018}) - uses $\sqrt{\text{SWAP}}$ gates to construct the Givens rotation.
  \item Controlled-$R_y$ method (Ref.~\cite{Xanadu_Quantum22}) - equivalent to a breakdown of controlled-$R_y$ followed by CNOT gates.
\end{enumerate}

At the raw circuit level, these methods require 2, 4, and 3 CNOT gates per Givens rotation, respectively.
However, we found the number of corresponding 2-qubit native gates for the transpiled/compiled circuits are equivalent to each other.
As a result, the total number of the 2-qubit gates for pUCCD(GS) and pUCCD(G) ansatze
scales as $3 N_G$ and $2 N_G$, respectively. Here, $N_G$ is the number of Givens rotations.
Thus, the choice of the implementation method can be made based on convenience and the first method was adopted in this work.

Quantum circuits submitted to the hardware are transformed to native gate sets of the adopted hardware.
On the RIKEN-Quantinuum's Reimei device, two-qubit native gates are \texttt{ZZPhase} gates,
$\exp(-i \pi \alpha /2 (Z_p \otimes Z_q))$.
Here we follow the notation of pytket~\cite{tket,*pytket-quantinuum}, i.e., angle parameter is given in the unit of $\pi/2$.
In practice, the compiled circuit uses exactly twice as many \texttt{ZZPhase} gates as there are Givens rotations.

\subsection{Hardware information \label{subsec:hardware_info}}

We now describe the quantum hardware used in this work.
The Reimei is based on Quantinuum's H-1 series trapped-ion quantum computer,
and has 20 qubits with all-to-all connectivity.
The Reimei device was installed in RIKEN's Wako campus, and started its operation in February 2025.
Table~\ref{tab:Reimei_status} summarizes the calibration data of the Reimei device, which were measured on 2025-05-09.
The reported metrics include single- and two-qubit gate errors,
SPAM (state preparation and measurement) errors, crosstalk, and memory error rates.
These parameters provide a baseline for evaluating the quality
of quantum circuits executed on the device.
In particular, the relatively low one- and two-qubit gate error rates,
combined with the all-to-all connectivity of trapped-ion hardware,
enable the accurate implementation of the pUCCD ansatz circuits described in this work.

The experiments on the Reimei trapped-ion quantum computer of this work were carried out via the Quantinuum Nexus~\cite{qnexus} platform,
which enables users to run quantum simulations on various backends, including noise-free statevector simulators, emulators, and real devices.

\begin{table}[hbtp]
  \begin{ruledtabular}
  \caption{
  Calibration data of the Reimei device.
  The values are from data on 2025-05-09 via
  the Quantinuum's Nexus~\cite{qnexus} platform.
  \label{tab:Reimei_status}}
    \begin{tabular}{lrr}
      Error type                 & Value               & Uncertainty         \\ \hline
      One-qubit gate error $P_1$ & $4.22\times10^{-5}$ & $5.23\times10^{-6}$ \\
      Two-qubit gate error $P_2$ & $1.39\times10^{-3}$ & $5.92\times10^{-5}$ \\
      SPAM error (0)             & $2.73\times10^{-3}$ & $1.57\times10^{-4}$ \\
      SPAM error (1)             & $5.25\times10^{-3}$ & $2.18\times10^{-4}$ \\
      Crosstalk error            & $3.21\times10^{-5}$ & $1.89\times10^{-6}$ \\
      Memory error               & $2.83\times10^{-4}$ & $2.32\times10^{-5}$ \\
   \end{tabular}
  \end{ruledtabular}
\end{table}

\begin{figure*}%[h]
  \centering
  \includegraphics[width=1.0\linewidth]{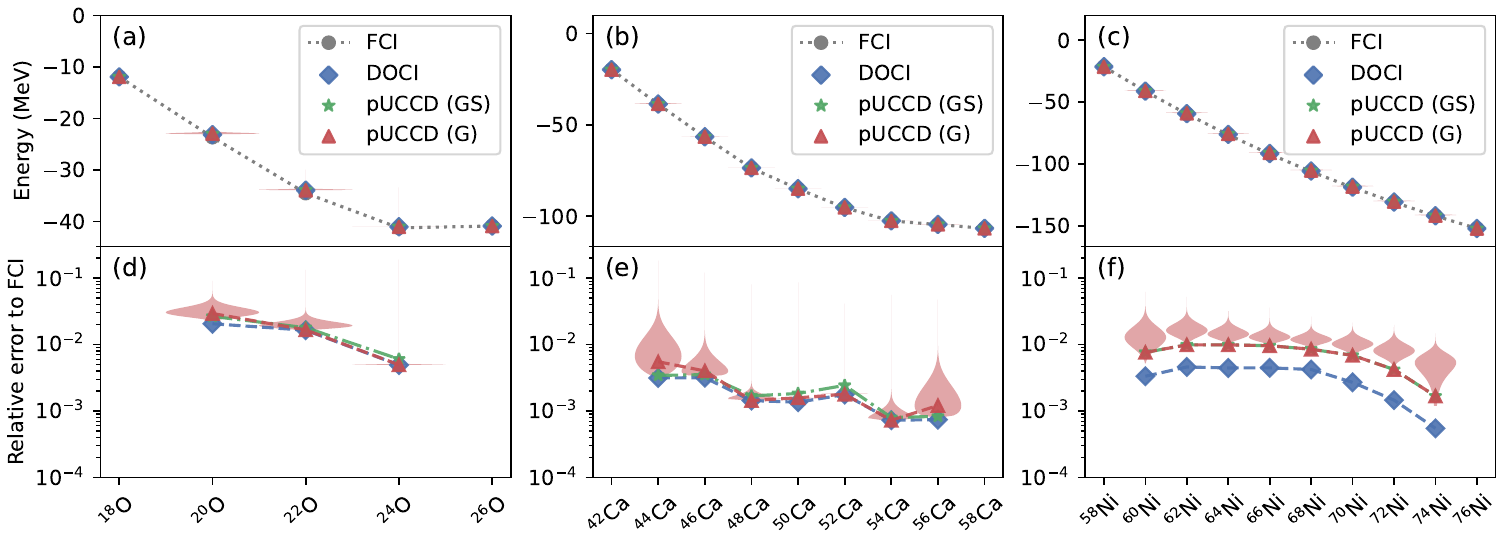}
  \caption{
    Ground-state energies of target isotopes on noise-free statevector simulators.
    (a)--(c) Ground-state energies of oxygen, calcium, and nickel isotopes, respectively.
    Full configuration interaction (FCI, dotted lines with gray circles) and 
    doubly occupied configuration interaction (DOCI, blue diamonds) are compared with statevector simulations using the pUCCD(GS) ansatz (green stars) and the pUCCD(G) ansatz (red triangles).
    %Here, pUCCD(GS) denotes the pUCCD ansatz implemented with nearest-neighbor Givens rotations and SWAP gates, while pUCCD(G) assumes all-to-all connectivity, as realized in the Reimei device.
    (d)--(f) Differences between the pUCCD ansatz results and the DOCI results are shown relative to FCI in logarithmic scale.    
    Uncertainty bands for pUCCD(G) are obtained by random sampling over circuits with different initial configurations and Givens-rotation orderings.
    \label{fig:FCI_DOCI_pUCCD}
  }
\end{figure*}

%
%%%%%%%%%%%%%%%%%%%%%%%%%%%%% 
%%%% +++++ Results +++++ %%%%
%%%%%%%%%%%%%%%%%%%%%%%%%%%%%   
%
\section{Results\label{sec:Results}}

In this section, we present the results of the pUCCD ansatz 
both on noise-free statevector simulators and on the Reimei quantum hardware
for the ground-state energies of oxygen, calcium, and nickel isotopes.
The former allows us to assess the intrinsic accuracy of the pUCCD ansatz,
while the latter demonstrates the feasibility of executing the pUCCD ansatz on real quantum hardware.

\subsection{Ideal simulation results of the pUCCD ansatz \label{subsec:Results_pUCCD}}

We first assess the performance of the pUCCD ansatz on a noiseless statevector simulator,
thereby probing its intrinsic accuracy in approximating ground-state energies of the target isotopes.
Full Configuration Interaction (FCI) provides
the exact ground-state energies within the valence space,
serving as a benchmark for evaluating the accuracy of approximate methods.
Doubly Occupied Configuration Interaction (DOCI) is another approximate method
that restricts the configuration space to zero-seniority configurations,
i.e., states in which all nucleons are paired as time-reversed partners.
In noise-free simulations, any deviation from FCI obtained with
the pUCCD ansatz reflects intrinsic limitations of the ansatz itself,
independent of hardware noise or statistical noise due to the finite 
number of measurement shots.

In what follows, we consider two types of pUCCD ansatz, named pUCCD(GS) and pUCCD(G).
The (GS) represents Givens plus SWAP, which consists of only nearest neighbor Givens rotations
and thereby more suitable for hardware with a limited connectivity~\cite{Google_2020,Elfving_PRA2021}.
On the other hand, (G) stands for Givens rotations where all-to-all connectivity is assumed.
In experiments on the Reimei device, we will focus on the pUCCD(G) ansatz.

Fig.~\ref{fig:FCI_DOCI_pUCCD} shows ground-state energies for the target isotopes,
obtained with pUCCD(G) and pUCCD(GS) on noise-free statevector simulators
compared to FCI and DOCI results.
The upper panels (a)--(c) display the ground-state energies.
At this scale, all curves are nearly indistinguishable.
The lower panels (d)--(f) show the differences between pUCCD and DOCI results
relative to FCI in logarithmic scale.
Results for two-neutron (or two-neutron-hole) systems are omitted in the lower panels
where both DOCI and pUCCD exactly reproduce FCI in these cases.
This is because the HCB representation with paired operators can e
xactly represent the ground states of single pair systems without approximation~\cite{SY_PRC2024}.

For the pUCCD ansatze, the symbols show the results of one representative realization of the ansatz,
and the uncertainty bands are plotted for the pUCCD(G) ansatz.
We generated many pUCCD(G) realizations by varying the initial configuration
and the order of Givens rotations in the ansatz circuit.
The resulting energies from a skewed distribution around the representative realization,
and the uncertainty bands are constructed by fitting Gamma distributions to the histograms of sampled energies.
More technical details are provided in the Supplemental Material~\cite{Supple}.

Most importantly, pUCCD reproduces DOCI energies with an error of mostly less than $1\%$, while using far fewer parameters, demonstrating its compact yet
accurate description. For reference, DOCI itself reproduces FCI ground-state energies within $0.1 - 1 \%$, with deviations smaller than the typical
uncertainties (a few percent) of modern nuclear many-body methods~\cite{HergertFP2020}.
The residual differences depend on the target system: Ca isotopes in
the $pf$ shell are well described within the zero-seniority space,
whereas oxygen isotopes show larger contributions from higher-seniority configurations
leading to slightly larger deviations from FCI.
Nickel isotopes in the $jj45$ shell exhibit larger discrepancies
between DOCI and pUCCD compared to oxygen and calcium isotopes.
This arises from the factors intrinsic to the $jj45$ shell.
The valence space of the $jj45$ shell consists of four $jj$-coupling orbitals ($1p_{1/2}, 1p_{3/2}, 0f_{5/2}, 0g_{9/2}$),
and the configuration mixing among these orbitals is pronounced in Ni isotopes even for ground states since their effective single particle energies are close to each other~\cite{JUN45}.
The results imply that the dominant pairing correlations in the same-species nucleons are expressed rather separable manner,
and that one may need to consider more expressive ansatze, e.g., by increasing the number of Trotter steps or including higher-order excitations,
to further improve the accuracy for systems with strong configuration mixing.

\begin{table*}%[htbp]%[!h]
\centering
\caption{Summary of pUCCD results on the RIKEN-Quantinuum Reimei device.
  FCI and ideal pUCCD are classical results.
  Reimei columns list results obtained on the Reimei device using two measurement strategies.
  ``Rel. Error'' columns show the absolute value of the relative error of the hardware results
  to the ideal pUCCD results, and the ``std.'' columns list typical 1$\sigma$ statistical
  uncertainties (in \%) from bootstrap analyses. Valid ratio is the fraction of bitstrings passing particle-number post-selection.
  All energies and their standard deviations are in MeV and relative errors are in \%.
  \label{tab:Reimei_results}
}
\begin{tabular*}{\textwidth}{@{\extracolsep{\fill}} 
    l
    S[table-format=-3.3]
    S[table-format=-3.3]
    S[table-format=-3.3]
    l
    S[table-format=1.2]
    S[table-format=-3.3]
    l
    S[table-format=1.2, input-symbols={<}, table-align-text-pre=false]
    S[table-format=1.2]}
    \toprule
    & {FCI} & {Ideal pUCCD} & \multicolumn{3}{c}{Reimei: Hadamard} & \multicolumn{3}{c}{Reimei: Basis rotation} & \\
     \cline{4-6} \cline{7-9}
     {Nuclide} & {Energy } & {Energy} & {Energy} & {std.} & {Rel. Error}  & {Energy} & {std.} & {Rel. Error } &  {Valid ratio (\%)} \\ 
    \hline
    ${}^{18}$O &   -11.932 &  -11.932 &  -11.802 & (0.209) & 1.09       &  -11.893 & (0.099) & 0.33       & 0.98 \\
    ${}^{20}$O &   -23.632 &  -22.939 &  -22.383 & (0.234) & 2.42       &  -22.854 & (0.140) & 0.37       & 0.98 \\
    ${}^{22}$O &   -34.498 &  -33.924 &  -33.560 & (0.222) & 1.07       &  -33.829 & (0.152) & 0.28       & 0.97 \\
    ${}^{24}$O &   -41.225 &  -41.021 &  -40.985 & (0.194) & 0.09       &  -41.027 & (0.150) & 0.02       & 0.97 \\
    ${}^{26}$O &   -40.869 &  -40.869 &  -40.613 & (0.176) & 0.63       &  -40.940 & (0.106) & 0.18       & 0.96 \\
    ${}^{42}$Ca &  -19.734 &  -19.734 &  -19.727 & (0.126) & 0.04       &  -19.775 & (0.053) & 0.21       & 0.97 \\
    ${}^{44}$Ca &  -38.675 &  -38.462 &  -38.540 & (0.162) & 0.20       &  -38.424 & (0.073) & 0.10       & 0.97 \\
    ${}^{46}$Ca &  -56.667 &  -56.440 &  -56.414 & (0.143) & 0.05       &  -56.565 & (0.062) & 0.22       & 0.96 \\
    ${}^{48}$Ca &  -73.662 &  -73.556 &  -73.538 & (0.131) & 0.02       &  -73.489 & (0.080) & 0.09       & 0.95 \\
    ${}^{50}$Ca &  -85.055 &  -84.921 &  -85.135 & (0.151) & 0.25       &  -85.007 & (0.087) & 0.10       & 0.94 \\
    ${}^{52}$Ca &  -95.360 &  -95.190 &  -95.038 & (0.146) & 0.16       &  -95.149 & (0.095) & 0.04       & 0.94 \\
    ${}^{54}$Ca & -102.632 & -102.557 & -102.622 & (0.127) & 0.06       & -102.648 & (0.081) & 0.09       & 0.94 \\
    ${}^{56}$Ca & -104.589 & -104.462 & -104.620 & (0.143) & 0.15       & -104.325 & (0.082) & 0.13       & 0.94 \\
    ${}^{58}$Ca & -106.666 & -106.666 & -106.885 & (0.132) & 0.20       & -106.668 & (0.058) & <0.01      & 0.94 \\
    ${}^{58}$Ni &  -21.447 &  -21.447 &  -21.207 & (0.149) & 1.12       &  -21.291 & (0.071) & 0.73       & 0.93 \\
    ${}^{60}$Ni &  -41.276 &  -40.961 &  -40.752 & (0.197) & 0.51       &  -40.992 & (0.094) & 0.07       & 0.94 \\
    ${}^{62}$Ni &  -59.602 &  -59.013 &  -58.703 & (0.228) & 0.52       &  -59.015 & (0.110) & <0.01      & 0.93 \\
    ${}^{64}$Ni &  -76.444 &  -75.687 &  -76.027 & (0.261) & 0.45       &  -75.765 & (0.116) & 0.10       & 0.92 \\
    ${}^{66}$Ni &  -91.941 &  -91.059 &  -90.552 & (0.245) & 0.56       &  -90.695 & (0.138) & 0.40       & 0.93 \\
    ${}^{68}$Ni & -106.140 & -105.237 & -105.169 & (0.263) & 0.06       & -105.288 & (0.133) & 0.05       & 0.92 \\
    ${}^{70}$Ni & -119.034 & -118.215 & -118.189 & (0.253) & 0.02       & -118.158 & (0.115) & 0.05       & 0.91 \\
    ${}^{72}$Ni & -130.930 & -130.379 & -130.511 & (0.223) & 0.10       & -130.397 & (0.092) & 0.01       & 0.95 \\
    ${}^{74}$Ni & -141.932 & -141.694 & -141.565 & (0.197) & 0.09       & -141.715 & (0.074) & 0.02       & 0.93 \\
    ${}^{76}$Ni & -152.103 & -152.103 & -151.986 & (0.133) & 0.08       & -152.054 & (0.053) & 0.03       & 0.94 \\
\hline
\hline
\end{tabular*}
\end{table*}

\subsection{Results on RIKEN-Quantinuum's Reimei trapped-ion quantum device \label{subsec:Reimei}}

\begin{figure}%[htbp]
  \centering
  \includegraphics[width=0.95\linewidth]{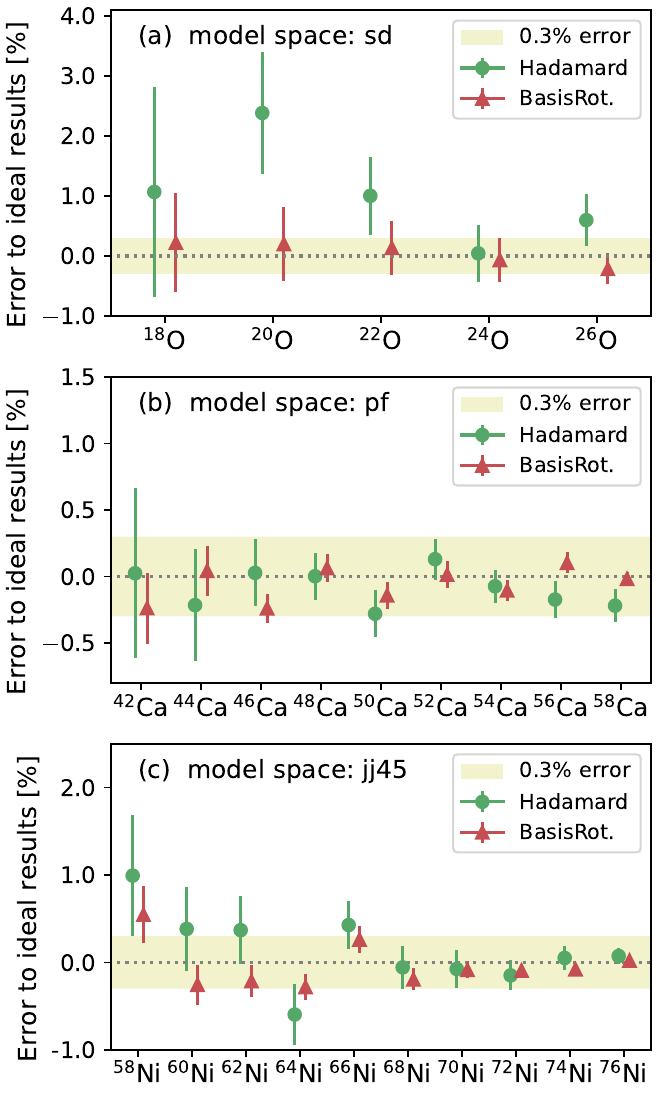}
  \caption{Relative errors in ground-state energies on the RIKEN-Quantinuum Reimei device.
  The results are shown for oxygen (a), calcium (b), and nickel (c) isotopes,
  using two measurement strategies: Hadamard and basis rotation.
  The error bars represent 1 $\sigma$ statistical uncertainties evaluated by bootstrapping,
  and the shaded band corresponds to a $0.3\%$ error.
  \label{fig:Reimei_results}
  }
\end{figure}

\begin{figure}%[h]
  \centering
  \includegraphics[width=0.95\linewidth]{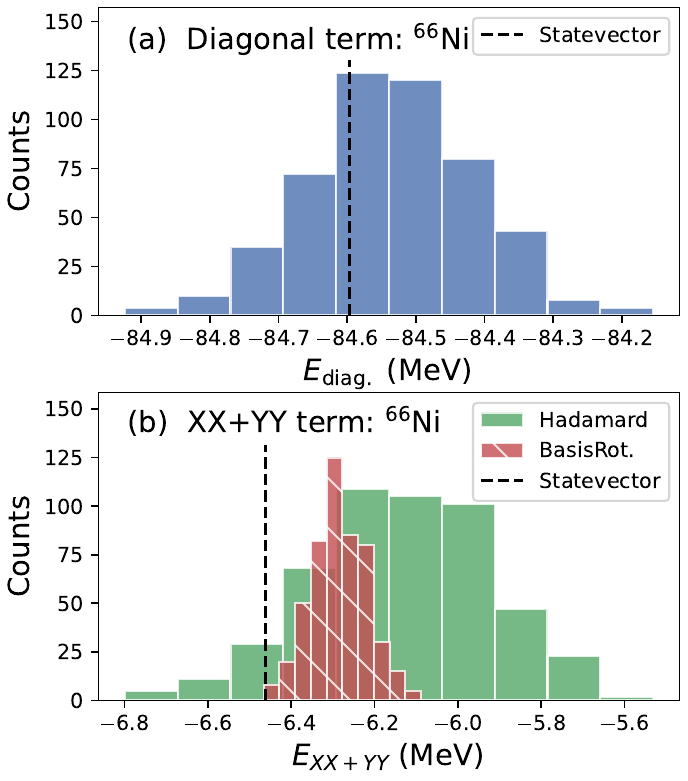}
  \caption{
    Bootstrapping of hardware results for ${}^{66}\mathrm{Ni}$.
    (a) diagonal terms. (b) $XX+YY$ terms evaluated by two measurement strategies: Hadamard (green) and basis rotation (red with hatching).
    The vertical dashed line shows the ideal statevector simulation result.
    The bin width is automatically determined by the Freedman-Diaconis rule.
    \label{fig:66Ni_bootstrap}
  }
\end{figure}

We now report the hardware results for ground-state energies of oxygen, calcium, and nickel isotopes
using the pUCCD(G) ansatz on Reimei.
Table~\ref{tab:Reimei_results} summarizes exact FCI energies,
the ideal pUCCD results from noise-free statevector simulations,
and the experimentally measured energies with two measurement strategies
along with percent errors relative to the ideal pUCCD results and valid ratios.
The valid ratio indicates the ratio of bitstrings
that preserve the correct particle number of the target nucleus after measurement,
reflecting the hardware fidelity in preparing the desired state.
The numbers are from ansatz measurements, hence independent of the measurement strategy for $XX+YY$ terms.

As a whole, the hardware results agree closely with the noise-free simulations,
showing relative errors mostly below $0.3\%$ to the ideal pUCCD results,
and at most a few percent relative to FCI.
These results demonstrate the capability of current quantum hardware to accurately capture nuclear pairing correlations.

This trend is further supported by bootstrapping analyses.
Fig.~\ref{fig:Reimei_results} displays the relative errors
of ground-state energies obtained on Reimei for oxygen, calcium, and nickel isotopes,
using both measurement strategies.
Error bars represent 1 $\sigma$ statistical uncertainties evaluated by bootstrapping
with 500 resamplings of datasets obtained from 1,024 measurement shots.
The shaded band corresponds to a $0.3\%$ error for reference.
Most data points fall within this range.

Note again that the difference between Hadamard and Basis rotation methods
lies only in the measurement of $XX+YY$ terms, while diagonal terms are measured in the same way for both methods.
Between the two measurement strategies, the basis-rotation method
consistently outperforms the Hadamard method, yielding results closer to ideal simulations.
Lighter isotopes exhibit larger uncertainties because
$XX+YY$ terms contribute to the total energy relatively larger in lighter isotopes.
In contrast, for heavier isotopes, the mass-dependent scaling commonly applied
to two-body matrix elements in phenomenological shell-model interactions
reduces the absolute size of the $XX+YY$ contributions
(e.g., TBMEs for ${}^{58}$Ca and ${}^{78}$Ni are multiplied by
$\sim 0.9$ compared with ${}^{42}$Ca and ${}^{58}$Ni),
which further diminishes their relative contribution to the total energy.

Next, we further examine the statistical uncertainties by separating the total energy into diagonal and $XX+YY$ contributions.
As an illustration, Fig.~\ref{fig:66Ni_bootstrap}, shows the distribution of the ground-state energy of ${}^{66}\mathrm{Ni}$
obtained by resampling the data 500 times.
The energy is separated into contributions from diagonal terms (top panel) and $XX+YY$ terms (bottom panel).
Vertical dashed lines indicate the corresponding noise-free statevector results for comparison.

The accuracy on the diagonal term depends primarily on occupation numbers,
while the $XX+YY$ terms are more sensitive to both occupation numbers
and relative phases between configurations.
As shown in the lower panel of Fig.~\ref{fig:66Ni_bootstrap}, the distribution obtained
using the Hadamard method is noticeably broader than that obtained with basis rotations.

We emphasize, however, that this analysis does not account for systematic errors
inherent to the Reimei device.
Therefore, our results do not imply that the basis-rotation method
always outperforms the Hadamard method in practice.
Nonetheless, the observed trends (see also Fig.~\ref{fig:Reimei_results})
suggest that basis rotations tend to yield more stable estimates of energy
expectation values for the systems studied here.

%%%%%%%%%%%%%%%%%%%%%%%%%%%%%%%%%%%%
%%%%%%% +++++ Summary +++++ %%%%%%%%
%%%%%%%%%%%%%%%%%%%%%%%%%%%%%%%%%%%%

\section{Summary and outlook \label{sec:Summary}}

Quantum many-body systems have become a central focus of
quantum computing research over the past decade,
driven by rapid progress in algorithms and hardware.
However, demonstrating genuine quantum advantage in interacting many-body physics remains an open challenge,
requiring advances in all components of the algorithmic workflow: ansatz design,
quantum algorithms, measurement strategies, optimization techniques, and so on.
This work concentrated on the ansatz and measurement aspects,
from the standpoint of nuclear structure, where exceptionally strong correlations
make the choice of ansatz especially critical.

By combining a hard-core boson mapping of the nuclear shell-model Hamiltonian 
with a pair-unitary coupled cluster doubles (pUCCD) ansatz, 
we demonstrated accurate ground-state calculations of oxygen, calcium, and nickel isotopes
on the RIKEN-Quantinuum Reimei trapped-ion quantum computer.
Our results achieve sub-percent agreement with noise-free statevector simulations,
setting a new benchmark for nuclear-structure simulations on real NISQ hardware.

While the pUCCD parameters in this work were optimized on classical simulators,
this proof-of-principle study shows that compact, physically motivated ansatze can
achieve high accuracy with modest qubit counts and circuit depths.
Such strategies can guide the design of efficient variational forms
for larger systems where classical pre-optimization is infeasible.
Optimization for classically intractable systems remains an open challenge,
and new, hardware-suitable optimization methods are needed.
Our previous study~\cite{SY_PRC2024} demonstrated that a sequential,
gradient-free optimization scheme~\cite{NFT_PRR2020} can effectively optimize a similar ansatz
for two-valence-neutron systems. Although those systems were still classically tractable,
the results suggest that sequential methods may be viable for optimizing or refining ansatze
on quantum hardware. For larger problems, hybrid approaches that combine classical pre-optimization
in a reduced subspace with subsequent quantum refinement could also be explored.

It remains an open question how best to extend the present framework to
include proton--neutron interactions, which are essential for a complete description of nuclei.
While several studies have investigated proton--neutron pairing
alongside like-particle pairing~\cite{np_PhysRevC.110.064320, ZhangPLB869_139841},
a practical quantum ansatz must be able to capture more general,
nonlocal correlations beyond simple pair excitations.
Developing such ansatze and evaluating their resource and measurement requirements will be necessary to treat mixed-species correlations robustly.
Nonetheless, our results demonstrate that pUCCD provides a compact and
accurate description of pairing correlations in single-species systems (multi-neutron or multi-proton),
establishing a foundation for future extensions to more complex nuclear systems,
and footsteps toward heavier systems such as neutron drops or neutron-rich isotopes.

As an illustration of possible extensions, entropy-driven entanglement
forging~\cite{PerezObiol2024} has recently been proposed to localize
correlations in Adapt-VQE circuits;
similar ideas including Ref.~\cite{PhysRevC.111.034317} could be combined with pUCCD,
for instance by monitoring generalized seniority.
In addition, hybrid qubit encoding schemes~\cite{delArcoSantos_2025} could be extended to nuclear systems.

Beyond ansatz design, new algorithmic paradigms merit exploration.
Quantum subspace diagonalization techniques~\cite{Stair2020,CortesPRA22,Kanno2023, RobledoMoreno2025, Yoshioka25},
including Quantum Krylov and related methods, use quantum devices to generate subspace states while delegating
diagonalization to classical post-processing.
These approaches provide access to excited states and transition properties such as electromagnetic transitions.
Such hybrid methods, in combination with physically motivated ansatze and developing efficient
schemes for time evolution and measurement, may offer a promising path toward realistic quantum simulations
of nuclear systems beyond the reach of classical methods.

\section*{Data Availability Statement}

The data presented in this manuscript,
including circuits and encoded Hamiltonians,
are available from the GitHub repository
and the Zenodo archive~\cite{DOI_Zenodo}.
For classical calculations, the exact diagonalization results can be
reproduced by using the NuclearToolkit.jl package~\cite{NuclearToolkit.jl, *Repo_NuclearToolkit.jl} with
the interaction files available through KSHELL~\cite{KSHELL1,*KSHELL2}.
We utilized the Qiskit~\cite{Qiskit}, PennyLane~\cite{PennyLane}, and pytket~\cite{tket,*pytket-quantinuum} for quantum computations.
Experiments on the Reimei device were carried out using the Quantinuum's Nexus~\cite{qnexus} platform.

%%%%%%%%%%%%%%%%%%%%%%%%%%%%%%%%%%%%%
%%%% +++++ Acknowledgments +++++ %%%%
%%%%%%%%%%%%%%%%%%%%%%%%%%%%%%%%%%%%%

\section*{acknowledgments}

This research was supported in part
by JSPS Grant-in-Aid for Scientific Research (Grant Nos.~JP22K14030, JP25H01511, JP25K01688),
JST PRESTO (Grant No.~JPMJPR25F8),
JST ERATO (Grant No.~JPMJER2304),
JST COI-NEXT (Grant No.~JPMJPF2221),
MEXT Q-LEAP (Grant No.~JPMXS0118067246),
and
RIKEN TRIP initiative (Nuclear Transmutation).
This work is based on the results obtained from a project, JPNP20017, commissioned by the New Energy and Industrial Technology Development Organization (NEDO).

%%%%%%%%%%%%%%%%%%%%%%%%%%%%%%%%%%%%%%%%%%%%%%%%%%%%%% 
\bibliography{ref} % for aps
%%%%%%%%%%%%%%%%%%%%%%%%%%%%%%%%%%%%%%%%%%%%%%%%%%%%%% 

 % %%%%%%%%%%%%%%%%%%%%%%%%%%%%%%%%%%%%%%%%%%
 % %%%%%%%%%%%%%%%%%%%%%%%%%%%%%%%%%%%%%%%%%%
 % %%%% +++++ Supplemental Matrial +++++ %%%%
 % %%%%%%%%%%%%%%%%%%%%%%%%%%%%%%%%%%%%%%%%%%
 % %%%%%%%%%%%%%%%%%%%%%%%%%%%%%%%%%%%%%%%%%%
 \clearpage
 \onecolumngrid
 \section*{Supplemental Material}

 In this Supplemental Material, we provide supplementary information regarding
 (i) the derivation of the hard-core-boson form of the nuclear Hamiltonian,
 (ii) the uncertainty bands of the pUCCD(G) ansatz,
 (iii) the results of the pUCCD(GS) ansatz on IBM's noisy simulator,
 and (iv) description of Zenodo archive containing data analyzed in this work.

 \section{Derivation of hard-core-boson form of the Hamiltonian \label{sec:app_Hamil}}

 Here we present the detailed derivation of the pair-wise Hamiltonians for the systems of interest,
 corresponding to Eq.~(2) of the main text.
 Starting from the fermionic Hamiltonian with up to two-body interactions,
 we introduced time-reversal pairs $(i, \bar{i})$ and defined pair creation, annihilation, and number operators
 in Eqs.~(4)--(5).

 After reorganizing terms and discarding those irrelevant within the zero-seniority space, the Hamiltonian reduces to the following forms:
 \begin{align}
   & \sum_{i} \epsilon_i a^\dagger_i a_i  = \sum_{i<0} ( \epsilon_i  a^\dagger_i a_i + \epsilon_{\bar{i}}  a^\dagger_{\bar{i}} a_{\bar{i}} ) 
  = \sum_{\tilde{i}}  \epsilon_{\tilde{i}} N_{\tilde{i}}, \\
   &\frac{1}{4} \sum_{ijkl} V_{ijkl} a^\dagger_i a^\dagger_j a_k a_l
    =  \frac{1}{4} \sum_{ij} V_{i\bar{i}j\bar{j}} a^\dagger_i a^\dagger_{\bar{i}} a_{\bar{j}} a_j
 + \frac{1}{4} \sum_{i, j\notin \{i, \bar{i}\}} V_{ijij} a^\dagger_i a^\dagger_j a_j a_i 
 +  (\text{irrelevant terms}), \\
 & \frac{1}{4}\sum_{i, j\in \{i, \bar{i}\}} V_{i\bar{i}j\bar{j}} a^\dagger_i a^\dagger_{\bar{i}} a_{\bar{j}} a_j
  = \frac{1}{4} \left( \sum_{i} V_{i\bar{i}i\bar{i}} a^\dagger_i a^\dagger_{\bar{i}} a_{\bar{i}} a_i +
 \sum_{i} V_{i\bar{i}\bar{i}i} a^\dagger_i a^\dagger_{\bar{i}} a_{i} a_{\bar{i}} \right) 
 = \frac{1}{2} \sum_{i} V_{i\bar{i}i\bar{i}} a^\dagger_i a^\dagger_{\bar{i}} a_{\bar{i}} a_i
 = \sum_{i<0} V_{i\bar{i}i\bar{i}} a^\dagger_i a^\dagger_{\bar{i}} a_{\bar{i}} a_i,\\
 &\frac{1}{4}\sum_{i, j \notin \{i, \bar{i}\}} V_{i\bar{i}j\bar{j}} a^\dagger_i a^\dagger_{\bar{i}} a_{\bar{j}} a_j 
  = \sum_{i<0} \sum_{j \notin \{i, \bar{i}\}<0} V_{i\bar{i}j\bar{j}} a^\dagger_i a^\dagger_{\bar{i}} a_{\bar{j}} a_j
 = \sum_{i \neq  j < 0} V_{i\bar{i}j\bar{j}} a^\dagger_i a^\dagger_{\bar{i}} a_{\bar{j}} a_j, \\
 &\frac{1}{4} \sum_{i, j\notin \{i, \bar{i}\}} V_{ijij} a^\dagger_i a^\dagger_j a_j a_i
 = \frac{1}{4} \sum_{i, j\notin \{i, \bar{i}\}} V_{ijij} n_i n_j
 \end{align}
 where the indices $i,j,k,l$ denote a fermionic single-particle states having $\{n, l, j, j_z, t_z\}$ quanta.
 The $ \tilde{i}, \tilde{j} $ represent pair-wise basis states obtained by folding
 single-particle states $ i $ and $ \bar{i} $ (the time-reversed partner).
 Rewriting these in terms of pair operators defined in Eqs.~(3)--(5) of the main text, we obtain
 \begin{align}
   \sum_{i<0} V_{i\bar{i}i\bar{i}} a^\dagger_i a^\dagger_{\bar{i}} a_{\bar{i}} a_i 
   & \rightarrow \sum_{\tilde{i}} V^{p}_{\tilde{i}\tilde{i}} A^\dagger_{\tilde{i}} A_{\tilde{i}}, \\
   \sum_{i \neq  j < 0} V_{i\bar{i}j\bar{j}} a^\dagger_i a^\dagger_{\bar{i}} a_{\bar{j}} a_j
   & \rightarrow \sum_{\tilde{i} \neq \tilde{j}} V^{p}_{\tilde{i}\tilde{j}} A^\dagger_{\tilde{i}} A_{\tilde{j}}, \\
   \frac{1}{4} \sum_{i, j\notin \{i, \bar{i}\}} V_{ijij} n_i n_j
   & \rightarrow \sum_{\tilde{i} \neq \tilde{j}} V^m_{\tilde{i}\tilde{j}} N_{\tilde{i}} N_{\tilde{j}},
 \end{align}
 where we put the superscripts $p$ and $m$ to distinguish pairing and monopole terms, respectively.

 \section{Uncertainty Bands of the pUCCD(G) ansatz \label{sec:app_Uncertainty}}

 \begin{figure}[ht]
   \centering
   \includegraphics[width=0.95\linewidth]{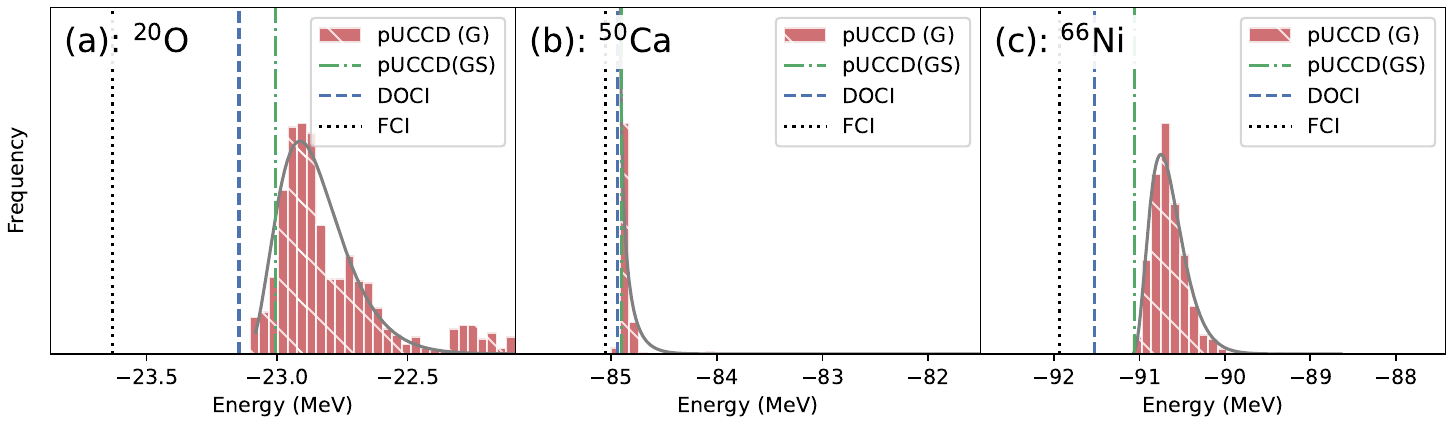}
   \caption{
     Uncertainty estimation for pUCCD(G) results.
     Sampled possible realization of ansatz is fitted by a Gamma distribution for the pUCCD results of (a) $^{20}$O, (b) $^{50}$Ca and (c) $^{66}$Ni.
     \label{fig:Random_pUCCD}
   }
 \end{figure}

 The pUCCD(G) ansatz can be realized in many different circuit layouts due to all-to-all connectivity. The number of possible circuits grows combinatorially with system size (e.g. for ${}^{50}$Ca, 252 possible reference states and 25! permutations of Givens rotations). Exhaustive sampling is thus impractical.

 To estimate uncertainties, we generated 1,000 random circuits per nucleus, varying both the initial occupied qubits and the ordering of Givens rotations. The resulting energy distributions (examples shown in Fig.~\ref{fig:Random_pUCCD} for ${}^{20}$O,
 ${}^{50}$Ca, and ${}^{66}$Ni) are asymmetric but well approximated by a Gamma distribution. To ensure physical plausibility, the lower bound of the distribution was set to the lowest energy encountered in the samples, thereby avoiding unphysical assignment of energies below DOCI.

 The fitted Gamma distributions were then used to construct the shaded uncertainty bands
 in Fig.~(3) of the main text. We note that the pUCCD(GS) ansatz typically yields 
 results close to the optimal edge of the pUCCD(G) distribution, 
 except for cases such as ${}^{24}$O, ${}^{50}$Ca, and ${}^{52}$Ca 
 where the qubit ordering in ascending order of total angular momentum $j$ leads
 to less favorable initialization.

 \section{Influence of depolarizing noise on pUCCD ansatz \label{sec:app_nonClifford}}

 \begin{table*}[htbp]
 \centering
 \caption{Relative errors of ground-state energies under depolarizing noise on single-qubit rotation gates
 with an error rate of $p=10^{-4}$. The ``cx/rx/ry/rz'' columns show transpiled gate counts, and the relative error is shown in percent.
 \label{tab:nonClifford_noise}}
 \begin{ruledtabular}
 \begin{tabular}{lrrrrr}
 Nucleus & cx & rz & ry & rx & Rel. error ($\%$) \\
 \hline
 $^{18}$O & 25  & 20  & 20  & 11  & 0.05 \\
 $^{20}$O & 40  & 32  & 32  & 18  & 0.08 \\
 $^{22}$O & 45  & 36  & 36  & 21  & 0.12 \\
 $^{24}$O & 40  & 32  & 32  & 20  & 0.11 \\
 $^{26}$O & 25  & 20  & 20  & 15  & 0.07 \\
 $^{42}$Ca & 45 & 36  & 36  & 19  & 0.03 \\
 $^{44}$Ca & 80 & 64  & 64  & 34  & 0.01 \\
 $^{46}$Ca & 105& 84  & 84  & 45  & 0.03 \\
 $^{48}$Ca & 120& 96  & 96  & 52  & 0.13 \\
 $^{50}$Ca & 125& 100 &100  & 55  & 0.16 \\
 $^{52}$Ca & 120& 96  & 96  & 54  & 0.10 \\
 $^{54}$Ca & 105& 84  & 84  & 49  & 0.19 \\
 $^{56}$Ca & 80 & 64  & 64  & 40  & 0.04 \\
 $^{58}$Ca & 45 & 36  & 36  & 27  & 0.03 \\
 $^{58}$Ni & 50 & 40  & 40  & 21  & 0.27 \\
 $^{60}$Ni & 90 & 72  & 72  & 38  & 0.17 \\
 $^{62}$Ni & 120& 96  & 96  & 51  & 0.19 \\
 $^{64}$Ni & 140&112  &112  & 60  & 0.23 \\
 $^{66}$Ni & 150&120  &120  & 65  & 0.12 \\
 $^{68}$Ni & 150&120  &120  & 66  & 0.13 \\
 $^{70}$Ni & 140&112  &112  & 63  & 0.18 \\
 $^{72}$Ni & 120& 96  & 96  & 56  & 0.06 \\
 $^{74}$Ni & 90 & 72  & 72  & 45  & 0.04 \\
 $^{76}$Ni & 50 & 40  & 40  & 30  & 0.03 \\
 \hline
 \end{tabular}
 \end{ruledtabular}
 \end{table*}

 In the FTQC era, it is expected that the non-Clifford gates will be
 the main bottleneck of quantum resource requirements
 due to the high cost of magic state distillation.
 However, in the NISQ era, one may focus on the noise effects
 on two-qubit gates such as CNOT and CZ gates,
 which are the main source of errors in current quantum hardware.
 To straddle the gap between the NISQ and FTQC era, we discuss the noise effects on
 the non-Clifford gates in the pUCCD ansatz.

 We consider a simple depolarizing noise model acting on each rotation gates in the pUCCD ansatz.
 The circuits are decomposed into single-qubit rotations and CNOT gates,
 and we assume that each CNOT gate is noise-free
 while each single-qubit rotation gate is followed by a depolarizing channel
 with an error rate $p=10^{-4}$.
 Under this model, we performed noisy simulations of the pUCCD(G) ansatz.
 In Table.~\ref{tab:nonClifford_noise}, we summarize the relative errors
 of ground-state energies for oxygen, calcium, and nickel isotopes
 with respect to the ideal statevector simulations.
 The results show that the depolarizing noise on single-qubit rotation gates
 induces only minor errors in the ground-state energies,
 with relative errors mostly below $0.01\%$.
 This suggests that the pUCCD ansatz is robust against such noise,
 and that the impact of non-Clifford gate errors may be limited in this context.
 One should note, however, that this conclusion is specific to the depolarizing noise model
 and the particular systems studied here with shallow circuits.
 For larger systems or different noise characteristics,
 the influence of non-Clifford gate errors may become more pronounced,
 necessitating further investigation.

 \newpage

 \section{Results of the pUCCD(GS) ansatz on IBM's noisy simulator}

 \begin{figure}[htbp]
   \centering
   \includegraphics[width=0.95\linewidth]{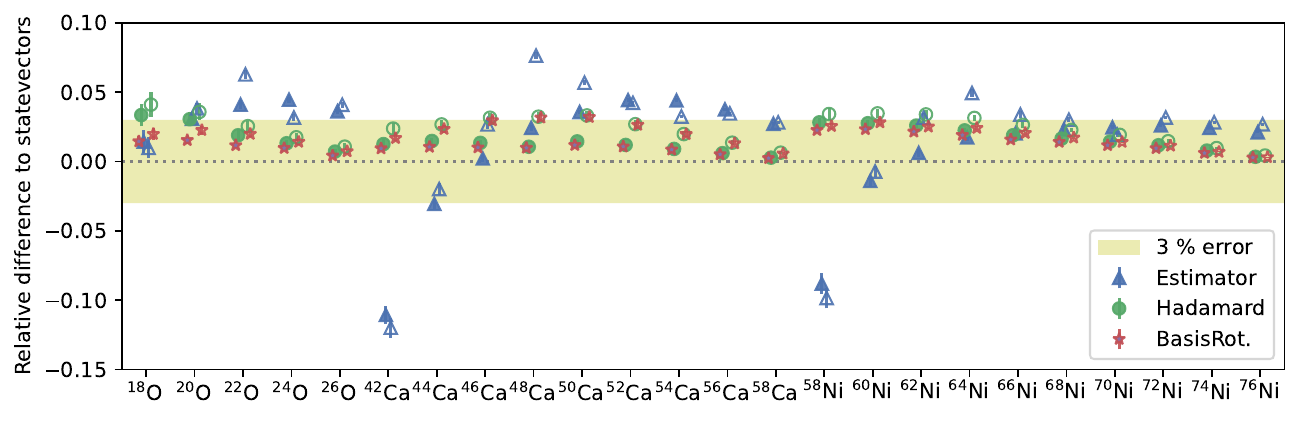}
   \caption{
 pUCCD results on IBM's noisy simulator.
 Relative error to the ideal 
 statevector simulation for the pUCCD ansatz. Closed markers correspond to 
 Givens+SWAP (GS) circuits, open markers to Givens-only (G) circuits. Colors and 
 markers indicate different measurement strategies as explained in the text. 
 The shaded band denotes $3\%$ deviation from the statevector results.
 Error bars show the standard deviation of 100 independent runs.
 \label{fig:simNISQvsSV}}
 \end{figure}

 For completeness, we benchmark the pUCCD(GS) ansatz on IBM’s noisy simulator 
 (FakeTorino backend), which reflects the heavy-hexagon connectivity of IBM devices.
 In this setting, additional SWAP gates are required, increasing the number of two-qubit gates to
 $3N_G$ (where $N_G$ is the number of Givens rotations), compared to $2N_G$ under 
 all-to-all connectivity.
 Figure~\ref{fig:simNISQvsSV} summarizes the relative errors with respect to the ideal 
 statevector simulations. Three measurement strategies were compared:
 \begin{itemize}
     \item \textbf{Estimator} (blue triangles): direct expectation-value evaluation.
     \item \textbf{Hadamard method} (green circles): Hadamard gates are applied to access 
     $XX$ terms, with $YY$ inferred by symmetry.
     \item \textbf{Basis-rotation method} (red stars): $XX+YY$ terms are diagonalized 
     via Givens rotations, enabling particle-number post-selection.    
 \end{itemize}

 While all approaches yield errors within a few percent, the basis-rotation 
 strategy consistently provides more stable results.
 In lighter nuclei, we observe a tendency to overestimate the binding energy with Estimator results.
 This behavior can be traced back to the measurement outcomes: bit strings corresponding to states with an incorrect particle number may occasionally appear.
 When such spurious configurations correspond to nuclei with a larger number of valence particles, they often yield lower expectation values of the Hamiltonian, i.e. artificial more bindings. As a result, the statistical averaging over these bit strings leads to an overbinding bias in light isotopes, where the relative contribution of such erroneous configurations is more pronounced.

 \section{Description of data archive \label{sec:app_DataArchive}}

 The data that support the findings of this study are openly available in:
 \begin{itemize}
     \item Zenodo at \url{https://doi.org/10.5281/zenodo.17217585} 
     \item GitHub repository at \url{https://github.com/SotaYoshida/Reimei_pUCCD_ShellModel}.
 \end{itemize}

 One can find the following data in the repository and archive:
 \begin{itemize}
     \item \texttt{circuit\_parameters}: Optimized parameters for pUCCD ansatze.
     \item \texttt{Data\_measured}: measured data on IBM's FakeTorino simulator and random sampling data for uncertainty estimation.
     \item \texttt{interaction\_file}: interaction files used in this work.
     \item \texttt{pytket\_circuits}: pytket circuit files for pUCCD ansatze.
     \item \texttt{qiskit\_Hamiltonians}: encoded Hamiltonians in qiskit format.
     \item \texttt{retrieved\_data}: measurement results retrieved from the Reimei quantum computer for both emulator and real hardware runs.
 \end{itemize}

\end{document}